\documentclass[twocolumn,showpacs,preprintnumbers]{revtex4-1}
\usepackage{graphicx}
\def \be{\begin{equation}}
\def \ee{\end{equation}}
\def \bdm{\begin{eqnarray}}
\def \edm{\end{eqnarray}}
\begin{document}
\setlength{\abovedisplayskip}{6pt}
\setlength{\belowdisplayskip}{6pt}
\preprint{Submitted to Physics of Plasmas}
\title{Time-Dependent Perpendicular Transport of Energetic Particles in Magnetic Turbulence with Transverse Complexity}
\author{A. Shalchi}
%
%
\affiliation{Department of Physics and Astronomy, University of Manitoba, Winnipeg, Manitoba R3T 2N2, Canada}
\date{\today}
\begin{abstract}
The motion of energetic particles in magnetic turbulence across a mean magnetic field is explored analytically.
The approach presented here allows for a full time-dependent description of the transport, including compound sub-diffusion.
The first time it is shown systematically that as soon as there is transverse structure of the turbulence, diffusion is restored
even if no Coulomb collisions are invoked. Criteria for sub-diffusion and normal Markovian diffusion are found as well.
\end{abstract}
\pacs{47.27.tb, 96.50.Ci, 96.50.Bh}
\maketitle
%
%
%
%
%
One of the most fundamental problems in plasma physics and astrophysics is to understand the motion of
energetic particles through a magnetized plasma. This motion is described by diffusion coefficients in the
different directions of space. In particular the transport of particles across a large scale or guide magnetic
field was subject of numerous analytical and numerical studies because of the complexity of this problem.
Knowledge of diffusion parameters is important for a variety of applications such as space weather studies,
acceleration of particles at shock waves, the propagation of cosmic rays through the universe, as well as
the motion of fast particles in a fusion reactor (see \cite{Schlickeiser2002,Wesson04,Balescu05}
for reviews). The simplest model for perpendicular transport is based on the assumption that particles follow magnetic
field lines while they move in the parallel direction with constant velocity. In this case the spread of particles
across the mean magnetic field is entirely controlled by the stochasticity of magnetic field lines. Analytically
this corresponds to the relation $\kappa_{\perp}=\kappa_{FL} v/2$ for the perpendicular diffusion coefficient
(see \cite{jok66}). Here we have used the particle speed $v$ and the field line diffusion coefficient $\kappa_{FL}$.
Characteristic for this model is that the corresponding perpendicular mean free path
$\lambda_{\perp}=3 \kappa_{\perp} /v=3 \kappa_{FL} /2$ does not depend on particle energy or momentum. In reality,
however, one expects that pitch-angle scattering influences the particle orbit and, therefore, the assumption of
an unperturbed motion is questionable. If strong pitch-angle scattering is present, perpendicular transport is
suppressed to a sub-diffusive level. This type of transport is usually called compound sub-diffusion
(see, e.g., \cite{KotaJokipii2000,Webbetal06}). Characteristic for this type of transport is that the mean square
displacement (MSD) of possible particle trajectories scales like $\langle ( \Delta x )^2 \rangle \propto \sqrt{t}$
with time. Therefore, normal Markovian diffusion, where we would have by definition $\langle ( \Delta x )^2 \rangle \propto t$,
cannot be found. Thus, the question arises what physical effect is required in order to restore normal diffusion.
According to the famous work of Rechester \& Rosenbluth (see \cite{rechrosen78}), Coulomb collisions can recover diffusion but cases
have been found in which lacking collisions entirely, perpendicular transport is still diffusive (see, e.g., \cite{giac99,qin02}).
In such numerical work evidence is provided that transverse complexity of the turbulence alone can restore Markovian diffusion,
at least in the late time limit. Obviously perpendicular transport is a complex non-linear process and, therefore,
it is challenging to develop analytical theories for this type of transport (see, e.g., \cite{rechrosen78,krom83,BAM97,matt03,shalchi04}).
The unified non-linear transport (UNLT, \cite{shal2010}) theory agrees with performed test-particle simulations for a variety
of magnetic field configurations including turbulence with small and large Kubo numbers as well as two-component turbulence
(see, e.g., \cite{taushal11,shalhus14}). However, such previous theories rely on the assumption of diffusive perpendicular transport.
Therefore, the following questions remain unanswered:

1. What exactly triggers perpendicular diffusion if there are no Coulomb collisions and what are the effects and
times scales leading to Markovian diffusion?

2. Can we develop a time-dependent theory of perpendicular transport which can describe compound sub-diffusion
at early times and then the restoration of diffusion for later times?

In the current letter we develop a time-dependent non-linear theory for perpendicular transport in order to answer
the those questions.

%
%
%
%
The fundamental equation describing the motion of charged particles through purely magnetic turbulence is the Newton-Lorentz
equation $d \vec{p} / d t = q \vec{v} \times \vec{B} (\vec{x}) / c$ where the total magnetic field is given as superposition
of a mean or guide field and a turbulent component, i.e. $\vec{B} (\vec{x}) = \delta \vec{B} (\vec{x}) + B_0 \vec{e}_z$. Furthermore,
we have used the electric charge of the energetic particle $q$, the particle velocity $\vec{v}$, the particle momentum $\vec{p}$,
as well as the speed of light $c$. If the particle position $\vec{x}$ is replaced by guiding center coordinates
$\vec{X} = \vec{x} + (\vec{v} \times \vec{e}_{z}) / \Omega$ (see, e.g., \cite{Schlickeiser2002}), the Newton-Lorentz
equation provides the following equations of motion
\be
V_x = v_z \frac{\delta B_x \left( \vec{x} \right)}{B_0} \quad \textnormal{and} \quad V_y = v_z \frac{\delta B_y \left( \vec{x} \right)}{B_0}
\label{eqmotion}
\ee
where $V_x$ and $V_y$ are the perpendicular components of the guiding center velocity. The gyrofrequency is given by
$\Omega=(q B_0)/(m c \gamma)$ with the rest mass of the particle $m$ and the Lorentz factor $\gamma$. In the following we
assume that the diffusion coefficients based on particle and guiding center coordinates are the same. In analytical
descriptions of turbulence, the magnetic field in Eq. (\ref{eqmotion}) is replaced by a Fourier representation. Problematic
in Eq. (\ref{eqmotion}) is the parallel component of the particle velocity $v_z$ because there is no simple way of modeling
this quantity due to the chaotic nature of the particle motion. Furthermore, it was shown that particle velocity and position
are strongly correlated (see, e.g., \cite{shal2010}). Based on \cite{shalchi05jgr} we write
\be
V_x (t) = \frac{1}{B_0} \int d^3 k \; \delta B_x \left( \vec{k} \right)
\frac{1}{i k_{\parallel}} \left( \frac{d}{d t} e^{i z k_{\parallel}} \right) e^{i \vec{x}_{\perp} \cdot \vec{k}_{\perp}}.
\label{eqmotion2}
\ee
A similar equation can be obtained for $V_y$. However, due to the assumption of axi-symmetric turbulence, this would
lead to the same result for the diffusion coefficient.
Perpendicular transport is described by the auto-correlation function $\langle V_x (t) V_x (0) \rangle$ where the guiding
center velocity is given by Eq. (\ref{eqmotion2}). In the following we assume that the magnetic fields and the phases in
Eq. (\ref{eqmotion2}) are uncorrelated. In the literature this type of approximation is either called {\it random phase
approximation} or {\it Corrsin's independence hypothesis} (see \cite{cor59}). This type of approximation could
be inaccurate in real magnetic turbulence which is spatially intermittent (see, e.g., \cite{burlaga04}). Furthermore,
we assume that the turbulence is homogeneous
$\langle \delta B_x (\vec{k}) \delta B_x^* (\vec{k}') \rangle = \delta (\vec{k}-\vec{k}') P_{xx} (\vec{k})$
where we have used the Dirac delta and the $xx$-component of the magnetic correlation tensor. As an additional
assumption we employ the hypothesis $\langle \exp{[i \vec{x} \cdot \vec{k}]} \rangle
\approx \langle \exp{[i z k_{\parallel}]} \rangle \langle \exp{[i \vec{x}_{\perp} \cdot \vec{k}_{\perp}]} \rangle$.
Therefore, we can write the auto-correlation function as
\be
\langle V_x (t) V_x (0) \rangle = \frac{1}{B_0^2} \int d^3 k \; P_{xx} \left( \vec{k} \right)
\xi \left( t, 0 \right) \left< e^{i \vec{x}_{\perp} \cdot \vec{k}_{\perp}} \right> 
\label{centraleq}
\ee
where we have used the parallel correlation function
\be
\xi \left( t_1, t_2 \right)
= k_{\parallel}^{-2} \left< \left( \frac{d}{d t_1} e^{i z (t_1) k_{\parallel}} \right) \left( \frac{d}{d t_2} e^{- i z (t_2) k_{\parallel}} \right) \right>
\label{definexi}
\ee
and assumed that $\vec{x}_{\perp} (t=0) = 0$. The parallel correlation function (\ref{definexi}) can be integrated
over times $t_1$ and $t_2$. Thereafter we calculate the derivatives with respect to $t_1$ and $t_2$. With $z(0)=0$ this trick
allows us to write
\be
\xi \left( t_1, t_2 \right)
= k_{\parallel}^{-2} \frac{d}{d t_1} \frac{d}{d t_2} \left< e^{i \left[ z \left( t_1 \right) - z \left( t_2 \right) \right] k_{\parallel}} \right>.
\label{finalxi}
\ee
In order to compute a time-dependent diffusion coefficient $d_{\perp} (t) = d \langle (\Delta x)^2 \rangle / (2 d t)$,
we employ the {\it TGK (Taylor-Green-Kubo) formulation} (see \cite{taylor22,green51,kubo57})
\be
d_{\perp} (t) = d_{\perp} (0) + \Re \int_{0}^{t} d t' \; \left< V_x (t') V_x (0) \right>
\label{tgk}
\ee
where we allow a non-vanishing initial diffusion coefficient. For the perpendicular characteristic function in
Eq. (\ref{centraleq}) we employ $\langle \exp{[i \vec{x}_{\perp} \cdot \vec{k}_{\perp}]} \rangle = \exp{[-\langle (\Delta x)^2 \rangle k_{\perp}^2 /2]}$ corresponding to a Gaussian distribution with vanishing mean. Furthermore, we can use Eqs. (\ref{centraleq})
and (\ref{tgk}) to write
\be
\frac{d^2}{d t^2} \langle (\Delta x)^2 \rangle = \frac{2}{B_0^2} \int d^3 k \; P_{xx} \left( \vec{k} \right)
\xi \left( t, 0 \right) e^{- \frac{1}{2} \langle (\Delta x)^2 \rangle k_{\perp}^2}.
\label{Generalsigma}
\ee
This ordinary differential equation can be evaluated for any given turbulence model described by the magnetic
correlation tensor $P_{nm}$ as long as the parallel correlation function $\xi (t_1, t_2)$ is specified as well.
It has to be emphasized that at not point we have assumed that perpendicular transport is diffusive.
In the following we combine Eq. (\ref{Generalsigma}) with a diffusion approximation and consider the case of weak and
strong pitch-angle scattering, respectively. Thereafter, we demonstrate how Eq. (\ref{Generalsigma}) explains compound
sub-diffusion and the recovery of diffusion for turbulence with transverse structure.

%
%
%
If pitch-angle scattering is suppressed we can set $z (t) = v \mu t$ in Eq. (\ref{finalxi}). Here we have used the
constant pitch-angle cosine $\mu$. For the perpendicular MSD we employ the diffusion approximation
$\langle (\Delta x)^2 \rangle = 2 D_{\perp} t$ where we have used the pitch-angle dependent Fokker-Planck
coefficient $D_{\perp} (\mu)$. Therewith, Eq. (\ref{Generalsigma}) becomes
\be
\langle V_x (t) V_x (0) \rangle
= \frac{v^2 \mu^2}{B_0^2} \int d^3 k \; P_{xx} \left( \vec{k} \right) e^{i v \mu t k_{\parallel} - D_{\perp} k_{\perp}^2 t}.
\label{QLT1}
\ee
Using Eq. (\ref{tgk}) with $d_{\perp} (0) = 0$, and integrating Eq. (\ref{QLT1}) over time gives
\be
D_{\perp} = \frac{v^2 \mu^2}{B_0^2} \int d^3 k \; P_{xx} \left( \vec{k} \right)
\frac{D_{\perp} k_{\perp}^2}{ \left( D_{\perp} k_{\perp}^2 \right)^2 + \left( v \mu k_{\parallel} \right)^2}.
\label{QLT2}
\ee
We can easily see that $D_{\perp} \propto |\mu|$. Therefore, we can write the solution as
$D_{\perp} = 2 \left| \mu \right| \kappa_{\perp}$ so that
$\kappa_{\perp} = \frac{1}{2} \int_{-1}^{+1} d \mu \; D_{\perp} (\mu)$
(see, e.g., \cite{Schlickeiser2002}). Thus, Eq. (\ref{QLT2}) becomes
\be
\kappa_{\perp} = \frac{v^2}{3 B_0^2} \int d^3 k \; 
\frac{P_{xx} (\vec{k})}{F (\vec{k}) + 4 \kappa_{\perp} k_{\perp}^2 / 3}
\label{QLT3}
\ee
with the function
\be
F (k_{\parallel},k_{\perp}) = (v k_{\parallel})^2 /(3 \kappa_{\perp} k_{\perp}^2).
\label{defF}
\ee
Eq. (\ref{QLT3}) can be written as $\kappa_{\perp} = v \kappa_{FL} /2$ where $\kappa_{FL}$ is the field line
diffusion coefficient. Combined with this form, Eq. (\ref{QLT3}) becomes equivalent to the integral equation
for field line diffusion derived in \cite{matt95}. The solution found here is called the field line
random walk limit. Eqs. (\ref{QLT2}) and (\ref{QLT3}) also contain quasi-linear theory originally derived
in \cite{jok66}. The latter theory can be obtained by employing the limiting processes $D_{\perp} \rightarrow 0$
or $\kappa_{\perp} \rightarrow 0$, respectively. In the opposite case one can find the result of Kadomtsev \& Pogutse
(see \cite{KadPog78}) as shown in \cite{shal2015}.

%
%
%
In astrophysics pitch-angle scattering is usually strong. Therefore, we now assume that parallel transport becomes diffusive
instantaneously and employ the characteristic function of a diffusion equation
$\langle \exp{[i \left( z (t_1) - z (t_2) \right) k_{\parallel}]} \rangle=\exp{[-\kappa_{\parallel} k_{\parallel}^2 |t_1-t_2|]}$.
Therewith, we derive from Eq. (\ref{finalxi})
\be
\xi \left( t_1, t_2 \right)
= - \kappa_{\parallel}^2 k_{\parallel}^2 e^{- \kappa_{\parallel} k_{\parallel}^2 \left| t_1 - t_2 \right|}.
\label{instantxi}
\ee
With the latter formula and the diffusion approximation $\langle (\Delta x)^2 \rangle = 2 \kappa_{\perp} t$,
Eq. (\ref{Generalsigma}) becomes
\be
\langle V_x (t) V_x (0) \rangle
= - \frac{\kappa_{\parallel}^2}{B_0^2} \int d^3 k \; P_{xx} ( \vec{k} ) k_{\parallel}^{2}
e^{- \kappa_{\parallel} k_{\parallel}^2 t - \kappa_{\perp} k_{\perp}^2 t}.
\label{strongvelocity}
\ee
The next step is the application of the {\it TGK formula} (\ref{tgk}). Here we have to be careful
because the running diffusion coefficient at initial time is not zero due to the assumption of instantaneous parallel
diffusion. Thus, we employ $d_{\perp} (t=0) = \kappa_{\parallel} \delta B_x^2 / B_0^2$ where we have used
$\delta B_x^2 = \int d^3 k \; P_{xx} (\vec{k})$. This corresponds to the assumption of instantaneous parallel diffusion
and the interaction with ballistic magnetic field lines (see e.g., \cite{ZybIsto85,shal2008}).
With $\kappa_{\perp} = d_{\perp} (t = \infty)$ we derive from Eq. (\ref{strongvelocity})
\be
\kappa_{\perp} = \frac{v^2}{3 B_0^2} \int d^3 k \; \frac{P_{xx} (\vec{k})}{F (\vec{k}) + v / \lambda_{\parallel}}
\label{StrongFinal}
\ee
with the parallel mean free path $\lambda_{\parallel} = 3 \kappa_{\parallel} / v$ and the function $F (k_{\parallel},k_{\perp})$
defined in Eq. (\ref{defF}). As demonstrated in \cite{shal2015}, Eq. (\ref{StrongFinal}) provides the same scaling as obtained
in the famous work of Rechester \& Rosenbluth if small Kubo number turbulence is considered. For the opposite
case the Zybin \& Istomin scaling (see \cite{ZybIsto85}) is obtained.
%
%
%

Eqs. (\ref{QLT3}) and (\ref{StrongFinal}) are very similar. In order to find an integral equation covering both cases,
we make the {\it Ansatz}
\be
\kappa_{\perp} = \frac{v^2}{3 B_0^2} \int d^3 k \; 
\frac{P_{xx} (\vec{k})}{F (\vec{k}) + (4/3) \kappa_{\perp} k_{\perp}^2 + v / \lambda_{\parallel}}.
\label{UNLT}
\ee
In the formal limit $\lambda_{\parallel} \rightarrow \infty$ we recover Eq. (\ref{QLT3}) and for $\lambda_{\parallel} \rightarrow 0$,
on the other hand, we find Eq. (\ref{StrongFinal}). Therefore, Eq. (\ref{UNLT}) correctly describes both cases, strong and
weak pitch-angle scattering. Eq. (\ref{UNLT}) is in perfect agreement with the integral equation provided by UNLT theory.
In \cite{shal2010} this theory was derived by employing lengthy calculations based on the pitch-angle dependent Fokker-Planck
equation. In the current paper we found an alternative but also a more intuitive derivation of UNLT theory.

%
%
%
In the following we drop the assumption of diffusive perpendicular transport but specify the properties of the magnetic fields.
As a first example we employ the slab model $P_{nm} (\vec{k})=g(k_{\parallel}) \delta (k_{\perp}) \delta_{nm} / k_{\perp}$ where
we have used the one-dimensional spectrum $g(k_{\parallel})$. For instantaneous parallel diffusion Eq. (\ref{centraleq}) becomes
\be
\langle V_x (t) V_x (0) \rangle
= - 4 \pi \frac{\kappa_{\parallel}^2}{B_0^2} \int_{0}^{\infty} d k_{\parallel} \; g \left( k_{\parallel} \right) k_{\parallel}^{2}
e^{- \kappa_{\parallel} k_{\parallel}^2 t}
\label{autocorr}
\ee
and from Eq. (\ref{tgk}) we derive for the running diffusion coefficient
\be
d_{\perp} (t) = 4 \pi \frac{\kappa_{\parallel}}{B_0^2} \int_{0}^{\infty} d k_{\parallel} \; g \left( k_{\parallel} \right)
e^{- \kappa_{\parallel} k_{\parallel}^2 t}.
\label{compoundform}
\ee
In the limit $t \rightarrow \infty$ we can approximate
\be
d_{\perp} (t) \approx \frac{4 \pi \kappa_{\parallel}}{B_0^2} g \left( k_{\parallel} = 0 \right)
\int_{0}^{\infty} d k_{\parallel} \; e^{- \kappa_{\parallel} k_{\parallel}^2 t}.
\label{doringdperp}
\ee
For the turbulence spectrum $g(k_{\parallel})$ we employ the Bieber et al. (see \cite{bieber94}) model
\be
g (k_{\parallel}) = \frac{1}{2 \pi} C(s) \delta B^2 l_{\parallel} \left[ 1 + (k_{\parallel} l_{\parallel})^2 \right]^{-s/2}
\label{slabspec}
\ee
with $C(s)=\Gamma (s/2)[2 \sqrt{\pi} \Gamma ((s-1)/2)]$ where we have used the inertial range spectral index $s$ and gamma functions. The parameter $l_{\parallel}$ is the correlation length in the parallel direction. Therewith, we derive from Eq. (\ref{doringdperp})
\be
d_{\perp} (t) = C(s) l_{\parallel} \frac{\delta B^2}{B_0^2} \sqrt{\frac{\pi \kappa_{\parallel}}{t}}
\label{compound}
\ee
corresponding to compound sub-diffusion as described before in \cite{KotaJokipii2000,Webbetal06}.

\begin{figure}
\includegraphics[width=.5\textwidth]{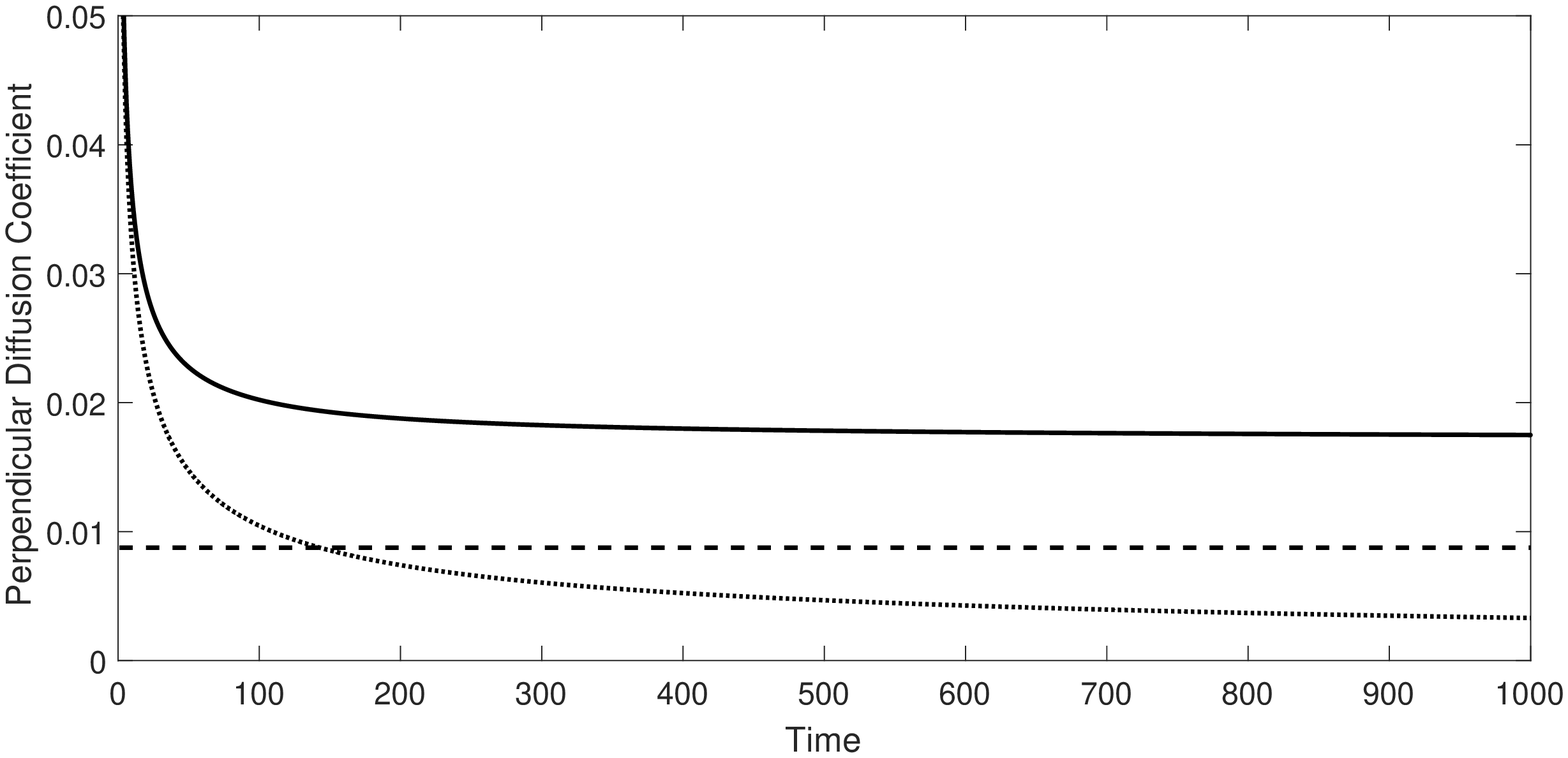}
\includegraphics[width=.5\textwidth]{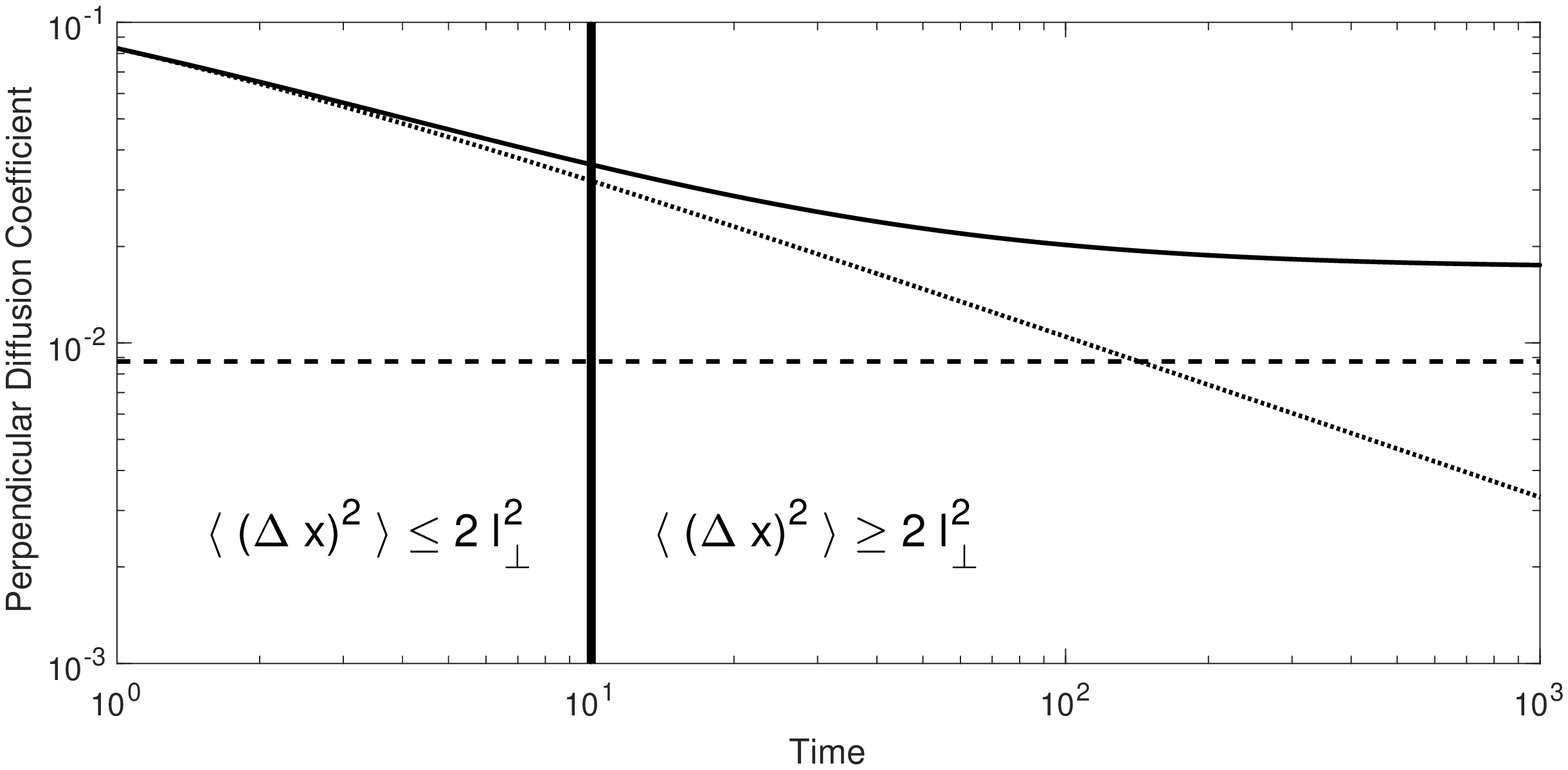}
\caption{The running diffusion coefficient $D_{\perp}$ versus time $\tau$ as obtained by solving
Eq. (\ref{Sigmanoisy3}) numerically for a Kubo number of $K=0.5$ (solid lines). The dotted lines describe
compound sub-diffusion as obtained from Eq. (\ref{compound}) and the dashed lines the (diffusive) CLRR limit
represented by Eq. (\ref{CLRR}). In the lower panel the graphs are shown as double-logarithmic plot to emphasize
the turnover from the sub-diffusive to the normal diffusive regime.}
\label{K05}
\end{figure}

\begin{figure}
\includegraphics[width=.5\textwidth]{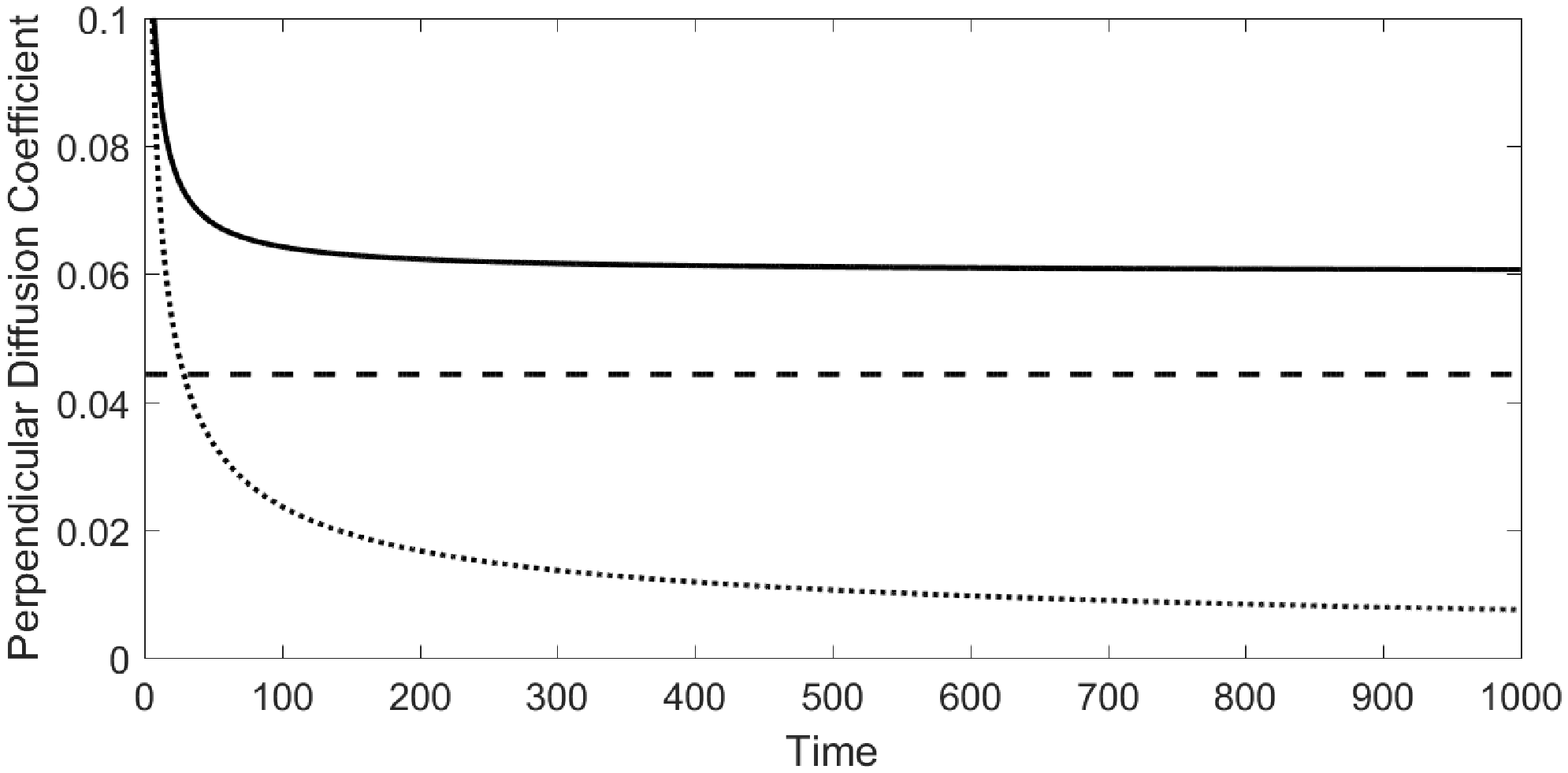}
\includegraphics[width=.5\textwidth]{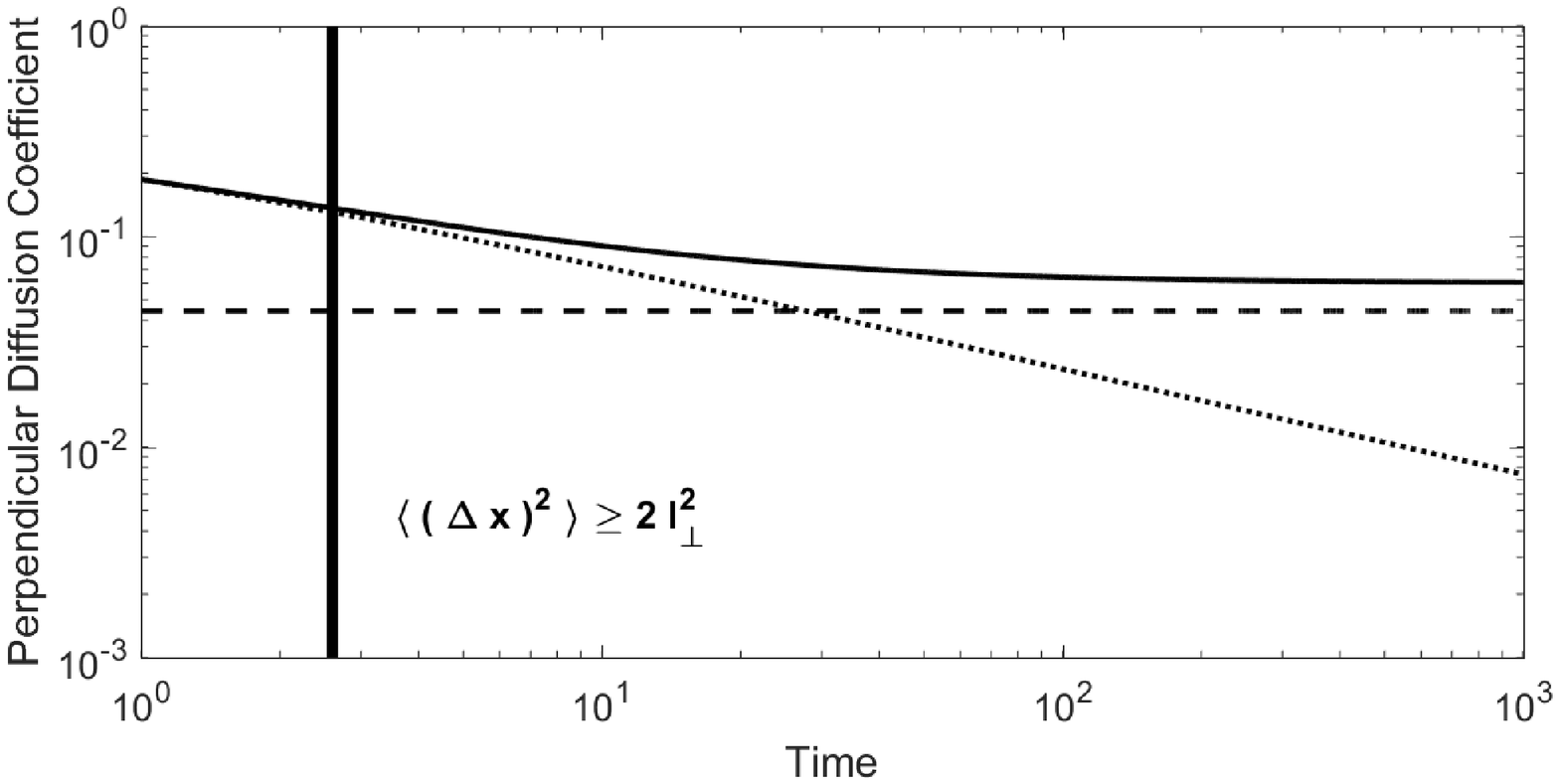}
\caption{As in Fig. \ref{K05}, but here we have considered a Kubo number of $K=0.75$.}
\label{K075}
\end{figure}

%
%
%
In order to restore diffusion in the collisionless case, we need transverse structure. Although more complex anisotropic
models for magnetohydrodynamic turbulence have been discussed in the literature (see, e.g., \cite{gs95,chan2000,yan02,Boldyrev06}),
we employ a simple noisy slab model defined via
\be
P_{mn} ( \vec{k} ) = \frac{2 l_{\perp}}{k_{\perp}} g (k_{\parallel}) \Theta \left( 1 - k_{\perp} l_{\perp} \right)
\left( \delta_{mn} - \frac{k_m k_n}{k_{\perp}^2} \right)
\label{noisyslab}
\ee
where we have used the {\it Heaviside step function} $\Theta (x)$ and the perpendicular correlation length of the
turbulence $l_{\perp}$. This model can be understood as broadened slab turbulence. If we assume again instantaneous
parallel diffusion, we can combine Eqs. (\ref{Generalsigma}), (\ref{instantxi}), and (\ref{noisyslab}).
The perpendicular wavenumber integral therein can be expressed by an {\it error function} and we derive
\bdm
\frac{d^2}{d t^2} \langle (\Delta x)^2 \rangle
& = & - 8 \pi \frac{\kappa_{\parallel}^2}{B_0^2} l_{\perp}
\sqrt{\frac{\pi}{2 \langle (\Delta x)^2 \rangle}} \textnormal{Erf} \left( \sqrt{\frac{\langle (\Delta x)^2 \rangle}{2 l_{\perp}^2}} \right) \nonumber\\
& \times & \int_{0}^{\infty} d k_{\parallel} \; g \left( k_{\parallel} \right) k_{\parallel}^{2}
e^{- \kappa_{\parallel} k_{\parallel}^2 t}.
\label{Sigmanoisy2}
\edm
One can easily demonstrate that for $\langle ( \Delta x )^2 \rangle \ll 2 l_{\perp}^2$ one recovers Eq. (\ref{autocorr})
and we find the sub-diffusive result discussed above. In order to find new physics, we need to satisfy
the condition $\langle ( \Delta x )^2 \rangle \gg 2 l_{\perp}^2$. For a numerical evaluation of Eq. (\ref{Sigmanoisy2}),
it is convenient to employ the integral transformation $x=l_{\parallel} k_{\parallel}$ and to use the Kubo number
$K = (l_{\parallel} \delta B_x)/(l_{\perp} B_0)$, the dimensionless time $\tau = \kappa_{\parallel} t / l_{\parallel}^2$,
as well as $\sigma = \langle (\Delta x)^2 \rangle / l_{\perp}^2$. With Eq. (\ref{slabspec}) for the
spectrum $g(k_{\parallel})$, we deduce
\bdm
\frac{d^2}{d \tau^2} \sigma
& = & - 4 \sqrt{\pi} C(s) K^2 \sqrt{\frac{2}{\sigma}} \;
\textnormal{Erf} \left( \sqrt{\frac{\sigma}{2}} \right) \nonumber\\
& \times & \int_{0}^{\infty} d x \; x^2 \left( 1 + x^2 \right)^{-s/2} e^{- \tau x^2}.
\label{Sigmanoisy3}
\edm
The latter differential equation can be solved numerically. The corresponding running diffusion coefficient
$D_{\perp}=(l_{\perp}^2 d_{\perp})/(l_{\parallel}^2 \kappa_{\parallel})$ is shown in Fig. \ref{K05} for
a Kubo number of $K=0.5$ and in Fig. \ref{K075} for a Kubo number of $K=0.75$, respectively.
We also show the {\it collisionless Rechester \& Rosenbluth (CLRR) limit}
\be
\frac{\kappa_{\perp}}{\kappa_{\parallel}}
= \left[ \frac{\pi}{2} C(s) \frac{l_{\parallel}}{l_{\perp}} \frac{\delta B^2}{B_0^2} \right]^2
\label{CLRR}
\ee
which can be derived from UNLT theory for small Kubo numbers and short parallel mean free paths (see \cite{shal2015}).
According to Figs. \ref{K05} and \ref{K075}, we find compound sub-diffusion for early times. As soon as the condition
$\langle ( \Delta x )^2 \rangle \gg 2 l_{\perp}^2$ is satisfied, normal Markovian diffusion is recovered due to the
transverse structure of the turbulence. The final diffusion coefficient is close to the CLRR limit.


In the current article we presented a detailed analytical study of perpendicular transport. The first time we have shown
analytically that perpendicular transport is diffusive even if there are no Coulomb collisions invoked. Perpendicular
diffusion is restored entirely due to transverse complexity of the turbulence. For turbulence without any transverse
structure we find the usual sub-diffusive behavior. As soon as the condition $\langle ( \Delta x )^2 \rangle \gg 2 l_{\perp}^2$
is satisfied, Markovian diffusion is recovered (see Figs. \ref{K05} and \ref{K075}). In combination with a diffusion
approximation, we found an alternative derivation of UNLT theory which showed good agreement with test-particle
simulations performed in the past. The original derivation of UNLT theory relies on lengthy calculations based on the
cosmic ray Fokker-Planck equation. The alternative derivation presented here, is shorter and more intuitive. The fact
that the same integral equation (see, e.g., Eq. (\ref{UNLT}) of the current paper) is obtained confirms the validity
of UNLT theory. Eq. (\ref{Generalsigma}) allows for a full time-dependent description of perpendicular transport.
Especially in small Kubo number turbulence, particles can move sub-diffusively for a long time before reaching the diffusive
regime. There will be several applications of the time-dependent description in a variety of physical scenarios ranging from
fast particles is fusion reactors to cosmic rays in the solar system, the interstellar medium, and other astrophysical systems.

Support by the Natural Sciences and Engineering Research Council of Canada (NSERC) is acknowledged.
{}

\end{document}